\documentclass[twocolumn]{aastex631}

\usepackage{savesym}
\usepackage{float}
\savesymbol{tablenum}
\usepackage[print-unity-mantissa=false, separate-uncertainty]{siunitx}
\DeclareSIUnit{\au}{AU}
\DeclareSIUnit\year{yr}
\DeclareSIUnit\gauss{G}

\restoresymbol{SIX}{tablenum}

\usepackage{amsmath}
\usepackage{amsfonts}
\usepackage{subfigure}
\usepackage{amssymb}

\begin{document}

\title{Stringent Upper Bounds on Atmospheric Mass Loss from Three Neptune-Sized Planets in the TOI-4010 System}

\correspondingauthor{Morgan Saidel}
\email{msaidel@caltech.edu}

\author[0000-0001-9518-9691]{Morgan Saidel}
\affiliation{Division of Geological and Planetary Sciences, California Institute of Technology, Pasadena, CA 91125, USA}

\author[0000-0003-2527-1475]{Shreyas Vissapragada}
\affiliation{Carnegie Science Observatories, 813 Santa Barbara Street, Pasadena, CA 91101, USA}

\author[0000-0002-0659-1783]{Michael Zhang}
\affiliation{Department of Astronomy \& Astrophysics, University of Chicago, Chicago, IL 60637, USA}

\author[0000-0002-5375-4725]{Heather A. Knutson}
\affiliation{Division of Geological and Planetary Sciences, California Institute of Technology, Pasadena, CA 91125, USA}

\author[0000-0001-6460-0759]{Matthäus Schulik}
\affiliation{Astrophysics group, Department of Physics, Imperial College London, Prince Consort Road, London SW7 2AZ, UK}

\author[0000-0002-1416-2188]{Jorge Fernández Fernández}
\affiliation{Department of Physics,
University of Warwick,
Coventry CV4 7AL
UK}

\author[0000-0001-9269-8060]{Michelle Kunimoto}
\affiliation{Department of Physics and Astronomy, University of British Columbia, 6224 Agricultural Road, Vancouver, BC V6T 1Z1, Canada}

\author[0000-0003-1452-2240]{Peter J. Wheatley}
\affiliation{Department of Physics, 
University of Warwick, 
Coventry CV4 7AL
UK}

\author[0000-0002-5547-3775]{Jessica Spake}
\affiliation{Carnegie Science Observatories, 813 Santa Barbara Street, Pasadena, CA 91101, USA}

\begin{abstract}

Photoevaporative models predict that the lower edge of the Neptune desert is sculpted by atmospheric mass loss. However, the stellar high energy fluxes that power hydrodynamic escape and set predicted mass loss rates can be uncertain by multiple orders of magnitude. These uncertainties can be bypassed by studying mass loss for planets within the same system, as they have effectively undergone scaled versions of the same irradiation history. 
The TOI-4010 system is an ideal test case for mass loss models, as it contains three Neptune-sized planets with planet b located in the `Neptune desert', planet c in the `Neptune ridge', and planet d in the `Neptune savanna'. 
Using Keck/NIRSPEC, we measured the metastable helium transit depths of all three planets in order to search for evidence of atmospheric escape. We place upper bounds on the excess helium absorption of 1.23\%, 0.81\%, and 0.87\% at 95\% confidence for TOI-4010~b, c and d respectively. We fit our transmission spectra with Parker wind models and find that this corresponds to 95th-percentile upper limits of $10^{10.17}$g~s$^{-1}$, $10^{10.53}$g~s$^{-1}$, and $10^{10.50}$g~s$^{-1}$ on the mass loss rates of TOI-4010~b, c, and d respectively. Our non-detections are inconsistent with expectations from one-dimensional hydrodynamic models for solar composition atmospheres. We consider potential reductions in signal from a decreased host star XUV luminosity, planetary magnetic fields, enhanced atmospheric metallicities, and fractionation, and explore the implications of our measurements for the past evaporation histories of all three planets.

\end{abstract}

\section{Introduction} \label{sec:intro}

\begin{figure}[h!]
    \centering
    \includegraphics[width=0.55\textwidth]{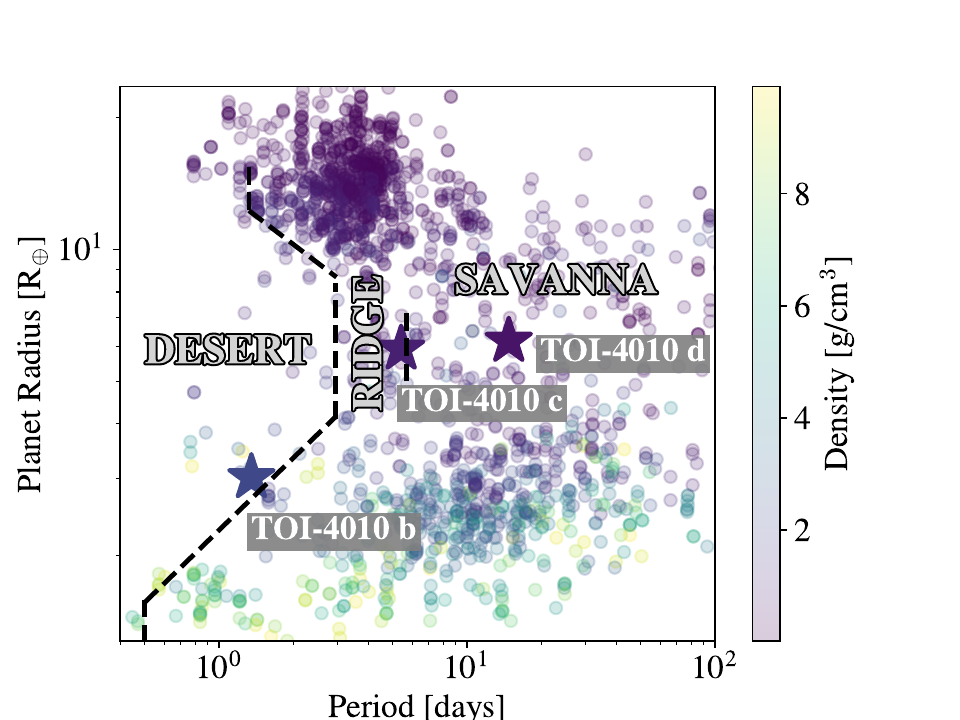}
    \caption{Transiting planet radii, periods, and densities in the vicinity of the Neptune desert, drawn from the NASA Exoplanet Archive on April 4, 2025 \citep{AkesonChen2013, ps}. The dashed lines indicate the Neptune desert, ridge, and savanna boundaries from \citet{Castro-GonzalezBourrier2024}. Planets in the TOI-4010 system are indicated with a star and a darker opacity.}
    \label{fig:TOI4010NeptuneDesert}
\end{figure}

Planets on close-in orbits can lose significant portions of their atmospheres to photoevaporative mass loss \citep[e.g.,][]{Owen2019}. 
It is thought that atmospheric mass loss may carve out part of the `Neptune desert' (see Figure~\ref{fig:TOI4010NeptuneDesert}), which is defined by a lack of sub-Neptune to Neptune-sized planets on close-in orbits \citep{SzaboKiss2011, BeaugeNesvorny2013, LundkvistKjeldsen2016}. Both theoretical \citep{OwenLai2018, IonovPavlyuchenkov2018} and observational work \citep{VissapragadaKnutson2022, GuilluyBourrier2023} suggest that the higher gravities of planets near the upper edge of the Neptune desert make them effectively invulnerable to photoevaporation.  As a result, it has been proposed that the shape of the upper edge is created by tidal disruption of planets that came too close to the star during high eccentricity migration \citep{OwenLai2018}.  The same theoretical models predict that the lower boundary of the desert is likely sculpted by atmospheric escape, and planets in this region are thought to have either formed in situ or migrated inward via interactions with the protoplanetary gas disk \citep{BatyginBodenheimer2016,BaileyBatygin2018,IdaLin2008}.

Our current story for the lower edge of the Neptune desert is based on predicted mass loss rates from population-level photoevaporative models, which can be uncertain by multiple orders of magnitude. These uncertainties come from poorly-constrained system properties that control mass loss, including the stellar Extreme Ultra-Violet (EUV) and X-ray (together XUV) radiation flux \citep[e.g.,][]{TuJohnstone2015}, stellar wind properties \citep{CarolanVidotto2020, MitaniNakatani2022, MacLeodOklopcic2022,RumenskikhKhodachenko2023}, the presence or absence of strong planetary magnetic fields \citep{SchreyerOwen2023}, and planetary atmospheric compositions \citep[e.g.,][]{ZhangKnutson2022}. In energy-limited mass loss models,  the total amount of mass lost is directly tied to the EUV and X-ray output of the host star. Most of this mass loss is thought to occur within the first $\sim$100 Myrs - 1 Gyr of the star’s lifetime, when its UV and X-ray output is highest \citep{Gudel2004,KingWheatley2018}. However, \citet{KingWheatley2021} found that significant EUV irradiation of exoplanet atmospheres can occur on Gyr timescales, indicating that EUV-driven mass loss is not necessarily dominated by the high-energy activity early in the star's lifetime, and can instead be important on Gyr timescales. Since different stars will have different time-integrated high energy fluxes, it can be difficult to compare
mass loss histories for planets in different systems. However, in a multi-planet system all planets experienced the same irradiation history scaled by orbital distance, effectively removing this as a source of uncertainty, although the overall uncertainty on the system-wide scaling of XUV fluxes remains \citep{OwenCamposEstrada2020}.

To date, there have been four published sets of mass loss measurements for multiple planets in the same system, including GJ 9827 b and d \citep{CarleoYoungblood2021}, HD 63433 b and c \citep{ZhangKnutson2022HD63433}, TOI-776 b and c \cite{LoydSchreyer2025}, and TOI-836 b and c \citep{ZhangBean2025}.  All of the planets in these systems have radii less than 3 R$_\earth$, and therefore host relatively low-mass (a few percent of the total planet mass at most) envelopes with a wide range of possible compositions. Atmospheric escape observations in these systems have typically focused on whether or not the planet can retain its primordial atmosphere.
These studies have found signatures of atmospheric escape from lower density mini-Neptunes \citep[$1.5-3$ R$_\earth$][]{ZhangKnutson2022HD63433, LoydSchreyer2025,ZhangBean2025}, and non-detections for super-Earths with higher bulk densities \citep[radii $\lesssim1.7$ R$_{\earth}$][]{CarleoYoungblood2021, ZhangKnutson2022HD63433,ZhangBean2025}.  These observations provide additional support for the hypothesis that atmospheric escape can strip the primordial hydrogen/helium envelopes of mini-Neptunes, leaving behind their rocky cores \citep[e.g.,][]{LopezFortney2013,OwenWu2017}.

In this study, we instead aim to use multi-planet systems to explore the origin of the Neptune desert. The TOI-4010 system \citep{KunimotoVanderburg2023} provides an ideal test case for this purpose, as this system consists of three Neptune-sized planets orbiting a K dwarf host. 
All three planets are located near the lower edge of the Neptune desert (see Figure~\ref{fig:TOI4010NeptuneDesert}), with TOI-4010~b (3.02 $R_{\oplus}$, $P = 1.35$ days) located inside the desert, TOI-4010~c (5.93 $R_{\oplus}$, $P = 5.41$ days) just outside the desert in the `Neptune ridge', and TOI-4010~d (6.18 $R_{\oplus}$, $P = 14.71$ days) in the middle of the `Neptune savanna' \citep[for definitions of these regions see][and also Figure~\ref{fig:TOI4010NeptuneDesert}]{Castro-GonzalezBourrier2024}. All three planets transit their stellar host, providing us with a unique opportunity to measure and compare mass loss rates from three planets located at different distances within the same system. The closely packed orbits of the three planets indicate that they cannot have migrated inward via high eccentricity channels, and are instead more analogous to the sub-Neptune-sized multi-planet systems along the lower edge of the desert.  The densities of these three planets (mass uncertainties are $<$12\%) show increasing envelope mass fractions with increasing distance \citep{KunimotoVanderburg2023}, and it is therefore particularly interesting to explore whether this gradient is primordial or could instead have been sculpted by mass loss processes. 

We can characterize the present-day properties of the atmospheric outflows from the TOI-4010 planets by measuring the absorption signature in the metastable helium feature (He$^*$) when each planet passes in front of its host star \citep{SpakeSing2018, OklopcicHirata2018}. Measuring He$^*$ outflow signals allows us to constrain the potential effects of varying atmospheric composition \citep{SalzCzesla2016, CrossfieldDragomir2020, DragomirCrossfield2020, ZhangKnutson2022}, the presence or absence of a strong planetary magnetic field \citep{TrammellArras2011, OwenAdams2014, SchreyerOwen2023}, and stellar winds \citep{ChristieArras2016, VidottoCleary2020, CarolanVidotto2021, WangDai2021} on the observed outflow properties of the three planets.  These mechanisms should result in distinct outflow morphologies, velocities, and overall mass loss rates \citep[e.g.,][]{WangDai2021,MacLeodOklopcic2022,SchreyerOwen2023}.

In this study, we use the Near Infrared Spectrograph (NIRSPEC) on Keck II Observatory to obtain He$^*$ transit observations of all three Neptune-sized planets in the TOI-4010 system. 
These observations allow us to expand our studies of atmospheric escape in multi-planet systems to larger (Neptune-sized) planets, which have massive hydrogen- and helium-dominated gas envelopes, for the first time. TOI-4010 is also the first three-planet system with mass loss constraints for all three planets. 
In Section~\ref{sec:datasection}, we detail our He$^*$ transit observations.  
In Section~\ref{sec:results}, we report the measured He$^*$ excess absorption signals from our Keck/NIRSPEC observations and in Section~\ref{sec:massloss} we model the corresponding mass loss rates for each planet. Lastly, in Section~\ref{sec:Discussion} we discuss the effect of varying stellar high-energy flux and atmospheric metallicity on the excess absorption measurements.

\section{Observations and Data Reduction} \label{sec:datasection}

Using Keck/NIRSPEC, we observed full transits of TOI-4010~b on October 13, 2023 UT, TOI-4010~c on January 6, 2024 UT, and TOI-4010~d on December 31, 2023 UT. All observations were obtained using the high-resolution mode and $Y$-band filter. Observations of TOI-4010~b and TOI-4010~d were obtained using the \ang{;;0.288}$\times$ \ang{;;12} slit, with an exposure time of \SI{300}{\second} for TOI-4010~b and an increased exposure time of \SI{600}{\second} for TOI-4010~d due to poor weather conditions which resulted in an increased PSF width and correspondingly lower flux through the slit. The seeing was even worse during our observations of TOI-4010~c, and as a result we used a wider \ang{;;0.432}$\times$ \ang{;;12} slit with an exposure time of \SI{600}{\second} in order to maximize flux. For all observations we used an ABBA nodding pattern to subtract the background. 

We analyzed the data using the pipeline described in \citet{ZhangKnutson2022, SaidelVissapragada2025}. Briefly, we identified bad pixels and produced a master flat, which we used to construct A-B subtracted images for each AB pair. We then used optimal extraction to obtain the 1D spectrum and corresponding uncertainties, and wavelength calibrated this spectrum using a PHOENIX model spectrum \citep{HusserWende-vonBerg2013} and model tellurics. Finally, we removed tellurics using \texttt{molecfit} \citep{SmetteSana2015} and interpolated all of the spectra onto a common wavelength grid.

\begin{figure*}[]
    \centering
    \includegraphics[]{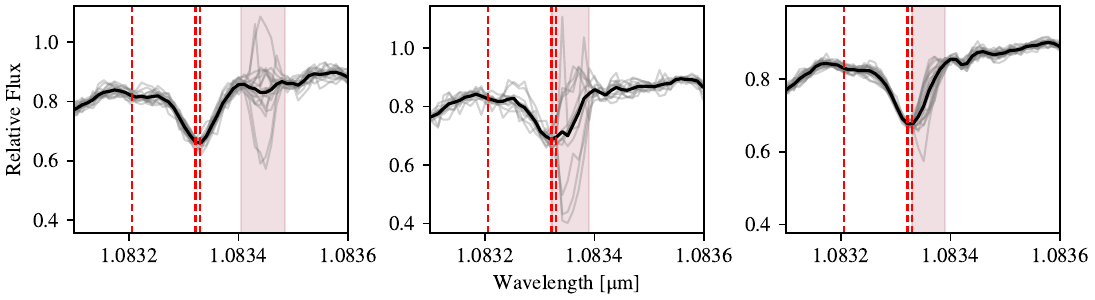}
    \caption{In-transit spectra of TOI-4010~b (first panel), TOI-4010~c (second panel), and TOI-4010~d (third panel) in the stellar rest frame. Gray lines are the in-transit spectra throughout the night. The average in-transit spectrum is shown in black. Red vertical dashed lines denote the wavelengths of the three helium lines. The red shaded regions correspond to the location of a telluric OH line and are masked in our analysis.}
    \label{fig:Tellurics}
\end{figure*}

Despite this correction, we found that the observations for all three planets were affected to varying degrees by residual telluric contamination (see Fig. \ref{fig:Tellurics}). Our observations of TOI-4010~b contain an OH telluric feature redward of the He$^*$ line, which is reasonably well-separated in wavelength space.  We have masked out this line (\SIrange[range-phrase=--,range-units=single]{1.08340}{1.08348}{\micro\meter}) as shown in the first panel in Figure~\ref{fig:Tellurics}. The location of the OH line shifts between observations, likely due to  changes in location of the stellar helium line between observations due to Earth's changing barycentric velocity, and we find that during our observations of TOI-4010~c and TOI-4010~d, the OH line now overlapped with the right side of the He$^*$ line. In order to ensure that the residual variations in this OH line do not affect our He$^*$ measurement, we have masked out the contaminated red wing of the line core (\SIrange[range-phrase=--,range-units=single]{1.08332}{1.08339}{\micro\meter}), as shown in Figure~\ref{fig:Tellurics} and focused our analysis only on the blue wing of the He$^*$ line.  

\begin{figure*}[]
    \centering
    \includegraphics[]{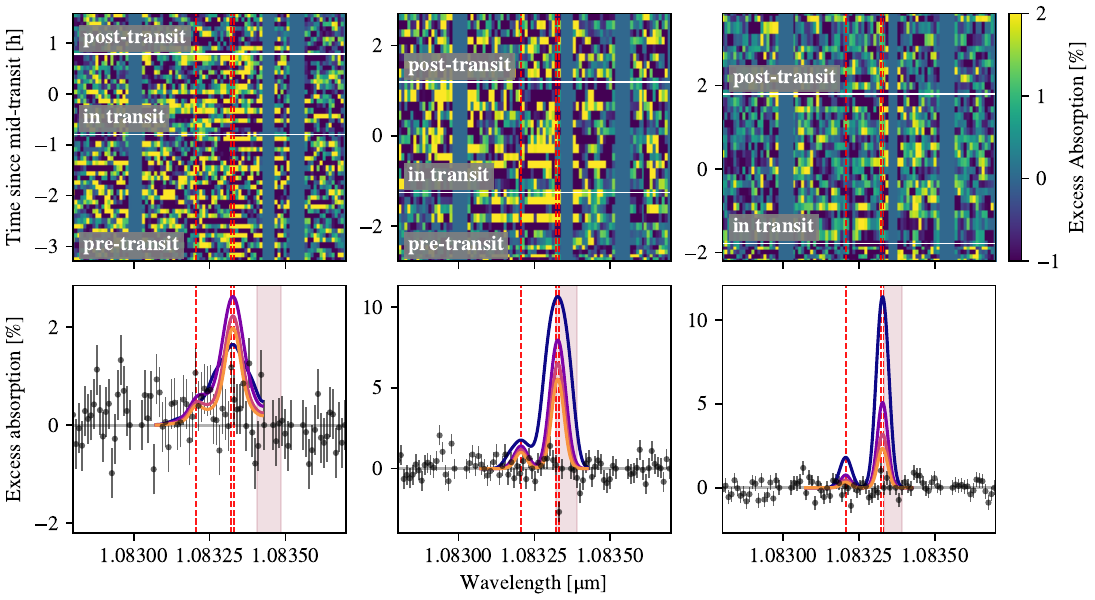}
    \caption{Top row: Keck/NIRSPEC excess absorption in percent of TOI-4010~b (first panel), TOI-4010~c (second panel) and TOI-4010~d (third panel), in each planet's rest frame as a function of time and wavelength (air wavelengths in planetary rest frame). Horizontal white lines mark the beginning (top) and end (bottom) of transit. Dashed vertical red lines denote the positions of the three helium lines. Red shaded regions mark the location of a telluric OH line that we masked in our analysis (see Fig.~\ref{fig:Tellurics}).
    Bottom row: Average excess absorption spectra for TOI-4010~b (first panel), TOI-4010~c (second panel) and TOI-4010~d (third panel) are shown as black points. The \texttt{pyTPCI} solar metallicity model predictions of the excess absorptions are shown as a blue line (nominal XUV flux), purple line (XUV flux reduced by factor of five), pink line (XUV flux reduced by factor of 10), and orange line (XUV flux reduced by factor of 15).}
    \label{fig:Keck_residuals}
\end{figure*}

Next, we isolated the time-varying component of the spectroscopic timeseries by taking the log of all telluric-corrected spectra on the common wavelength grid and subtracting the average of every row and column. The resulting timeseries of residual spectra contain the fractional flux variations, where each pixel corresponds to a specific wavelength and time. We then constructed our in-transit excess absorption spectrum by taking all columns (corresponding to wavelength) of the residuals image and subtracting off the average out-of-transit residual (defined as the residuals preceding first contact and following fourth contact).  Because the residuals are in log space, this is equivalent to dividing by the geometric mean. We removed continuum variations by masking out strong lines in the spectrum and fitting a third-order polynomial to the spectrum, then subtracting it from the residuals. We then computed the radial velocity of the planet relative to the star using each epoch's barycentric Julian date and shifted the spectrum onto the planetary rest frame. The resulting excess absorption spectrum for each planet is shown in Figure~\ref{fig:Keck_residuals}.

\section{Results}\label{sec:results}

All three planets display no discernible excess He$^*$ absorption. To set an upper limit to the possible He$^*$ absorption for each planet, we take the mean excess absorption within a \SI{0.75}{\angstrom} bin centered on the main helium triplet peak (\SI{10833.327}{\angstrom}). Following e.g. \citet{VissapragadaMcCreery2024}, we estimate $\sigma$ as the rms scatter of the transmission spectrum binned to \SI{0.75}{\angstrom}. For TOI-4010~b, we rule out absorption to $<$~1.23\% at 95\% confidence, for TOI-4010~c and TOI-4010~d we rule out absorption to $<$~0.81\% and $<$~0.87\% respectively at 95\% confidence.

\section{Mass Loss Modeling}\label{sec:massloss}

\subsection{Joint Constraints on Outflow Temperature and Mass Loss Rate from Retrievals}\label{sec:masslosssub}

Using our non-detections of He$^*$, we are able to place an upper limit on the mass loss rate for all three planets. We calculated this upper limit assuming an approximately solar ratio of hydrogen to helium \citep[90\% hydrogen, 10\% helium by number, e.g.,][]{ASCHWANDEN2014235} in a one-dimensional isothermal Parker wind model. This is the most commonly utilized model in the literature \citep[e.g.,][]{OklopcicHirata2018, DosSantosVidotto2022, LinssenShih2024}, making it easier to compare our results to those of other published studies.
 We used \texttt{p-winds} \citep{DosSantosVidotto2022} to calculate the expected helium excess absorption spectrum as a function of the assumed mass loss rate ($\dot{M}$) and outflow temperature ($T_0$). When calculating these models, we adopted the planetary and stellar parameters from \cite{KunimotoVanderburg2023}.  TOI-4010 does not have a measured high-energy stellar spectrum, and we therefore used the MUSCLES spectrum of HD~85512 \citep{YoungbloodFrance2017} as a proxy as the star is a close match in spectral type (both K stars), effective temperature (HD 85512: \SI{4715}{\kelvin}; TOI-4010: \SI{4960}{\kelvin}), surface gravity \citep[HD 85512: $10^{4.39}$~cm/s$^2$; TOI-4010: $10^{4.54}$~cm/s$^2$;][] {YoungbloodFrance2017,KunimotoVanderburg2023},  stellar rotation period  \citep[HD 88512: $\approx$~44-58 days; TOI-4010: 37.7 days;][]{LovisDumusque2011, PepeLovis2011, LaliotisBurt2023, KunimotoVanderburg2023}, and log~$R'_{\mathrm{HK}}$ \citep[HD 85512: -4.98; TOI-4010: -4.90;][]{CostesWatson2021,KunimotoVanderburg2023} to TOI-4010.

\begin{figure*}[]
    \centering
    \includegraphics[]{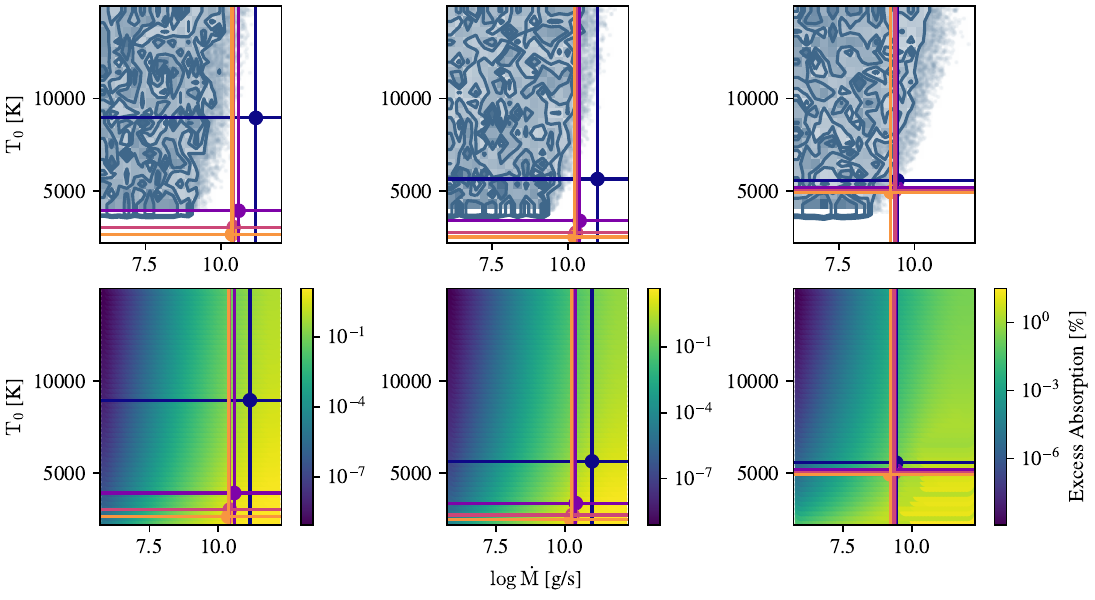}
    \caption{Top row: Posterior probability distribution for log $\dot{M}$ and $T_0$ for TOI-4010~b (first panel), TOI-4010~c (second panel), and TOI-4010~d (third panel). The contours indicate the 1$\sigma$, 2$\sigma$ and 3$\sigma$ levels for the distributions. The \texttt{pyTPCI} solar metallicity-predicted mass loss rates and estimated outflow temperatures are shown as blue circles (nominal XUV flux of stellar spectrum), purple circles (XUV flux reduced by factor of five), pink circles (XUV flux reduced by factor of ten), orange circles (XUV flux reduced by a factor of 15). Bottom row: \texttt{p-winds} predicted excess absorptions corresponding to our MCMC analysis for a $50\times50$ grid sampling our $\dot{M}$-$T_0$ prior range. 
    }
    \label{fig:MCMC}
\end{figure*}

We allowed $\dot{M}$ and $T_0$ to vary as free parameters and fit them to our data for each planet in order to obtain corresponding upper bounds on their outflow properties in $\dot{M}$-$T_0$ space.  We defined prior ranges of $\mathcal{U}(6, 12)$ for log $\dot{M}$ in \SI{}{g.s^{-1}} and $\mathcal{U}(3900, 15000)$ for $T_0$ in Kelvin, based on predicted isothermal outflow temperatures for sub-Neptune and Neptune-sized planets \citep[e.g.,][]{SalzCzesla2016,LinssenOklopcic2022} and used \texttt{emcee} to perform an MCMC analysis in order to retrieve the posterior probability distributions for log $\dot{M}$ and $T_0$. We verify that convergence is achieved by first confirming that the number of steps in our chain is greater than fifty times the autocorrelation length, indicating 
that we are running our chains for enough steps to generate independent samples.  We confirm convergence by verifying that the Gelman-Rubin statistic $\hat{R}$ is less than 1.01 for all sampled parameters.
The resulting constraints on the outflow properties of each planet as well as the predicted excess absorptions corresponding to a 50$\times$50 grid sampling our $\dot{M}$-$T_0$ prior range are shown in Figure~\ref{fig:MCMC}.

We find a 95th-percentile upper limit on log $\dot{M}$ of $10^{10.17}$g~s$^{-1}$ for TOI-4010~b, $10^{10.53}$g~s$^{-1}$ for TOI-4010~c and $10^{10.50}$g~s$^{-1}$ for TOI-4010~d. This is likely an over-estimate, as mass loss rates near these upper bounds would require unphysically high outflow temperatures (see discussion in \S\ref{sec:ares_models}).  Nonetheless, if we assume a constant mass loss rate equal to the stated upper limit for each planet, we estimate atmospheric lifetimes ($M_{p}/\dot{M}$) of $>$~\SI{110}{\giga\year} for all three planets (we note that this is a lower limit on the planet's lifetime as the star's XUV flux should decrease with age).  We conclude that even with this relatively conservative limit, photoevaporation will not significantly alter their atmospheric masses over the main sequence lifetime of the host star.

\subsection{Comparison to Predictions from Forward Models}\label{sec:ares_models}

Next, we consider whether or not these upper bounds are consistent with predictions from self-consistent forward models for solar metallicity atmospheres.  We calculated the predicted excess absorptions for each planet using \texttt{pyTPCI} \citep{RosenerZhang2025}, a self-consistent one-dimensional radiative-hydrodynamics code to model atmospheric outflows. 
Following the methodology in \cite{ZhangKnutson2022,ZhangBean2025}, we
set a lower atmospheric boundary condition at each planet's radii, defined a particle number density of \SI{1e14}{\centi\meter^{-3}}, and computed the corresponding atmospheric pressure using each planet's equilibrium temperature. As in \citet{ZhangBean2025}, we did not include molecules but did include atomic species with solar abundances greater than $10^{-5}$ (C, N, O, Ne, Mg, Si, S, and Fe), along with hydrogen and helium. We used each planet's mass and semi-major axis ($M_p$, $a$), and the same stellar spectrum of HD~85512 \citep{YoungbloodFrance2017} as a proxy for the unknown stellar spectrum of TOI-4010. We then modeled the predicted excess absorption signals and corresponding mass loss rates for each planet.  In order to facilitate comparisons with the isothermal outflow models in our MCMC retrievals, we identified the maximum value of the radially varying outflow temperature and 
used this value when comparing the retrieved versus predicted outflow properties in Fig. \ref{fig:MCMC}. 

We find that the predicted excess absorption signals for planets c and d should have been readily detectable in our observations, with a somewhat weaker predicted detection significance for planet b (see Fig. \ref{fig:Keck_residuals}) likely due in part to its much smaller radius and higher model-predicted outflow temperature ($T_0$ TOI-4010~b: \SI{8900}{\kelvin}; $T_0$ TOI-4010~c: \SI{5600}{\kelvin}; $T_0$ TOI-4010~d: \SI{5500}{\kelvin}).  The predicted outflow temperatures for all three planets lie near the lower end of the range considered in our retrievals.  When we compare the predicted outflow properties in $\dot{M}$-$T_0$ space to the constraints from our \texttt{p-winds} retrievals, we find that our upper bounds on the mass loss rates for planets b and c are approximately an order of magnitude lower than the predicted mass loss rates from the \texttt{pyTPCI} models ($\dot{M}$ for TOI-4010~b: $10^{11.15}$g~s$^{-1}$; $\dot{M}$ for TOI-4010~c: $10^{10.98}$g~s$^{-1}$) with the same outflow temperatures (see Fig. \ref{fig:MCMC}).  For planet d, our upper bound is consistent with the predicted mass loss rate of $10^{9.96}$g~s$^{-1}$.

\section{Discussion}\label{sec:Discussion}

There are several factors that could potentially explain our non-detections, including a lower than expected XUV luminosity for the host star, confinement and/or suppression of the outflow by stellar winds or planetary magnetic fields, an enhanced atmospheric metallicity, or fractionation in the outflow. We investigate each of these explanations in turn in the following sections as well as the past mass loss histories of these planets.

\subsection{Decreased Stellar XUV Flux}

Photoevaporative outflows are powered by high-energy radiation from the star.  If the true XUV flux of the host star was lower than that of our chosen stellar proxy (HD 85512), we would over-predict the mass loss rates and He$^*$ excess absorption signals for all three planets (we do note that stellar activity indicators such as the stellar rotation period and log~$R'_{\mathrm{HK}}$ are in good agreement for HD~85512 and TOI-4010 as discussed in Section~\ref{sec:masslosssub}, and the XUV flux of HD 85512 is in good agreement with the predicted present-day XUV flux of TOI-4010, see Section~\ref{sec:evo}). LTT 9779 b is a useful case in point, as \textit{XMM-Newton} observations revealed that the X-ray luminosity of LTT 9779 is 15$\times$ lower than expected for its age, drastically reducing its predicted mass loss rate \citep{FernandezFernandezWheatley2024}. Changes in the XUV flux can also result in a net change in the fractional abundance of He$^*$ in the outflow, as He$^*$ is populated via the ionization and subsequent recombination of helium atoms \citep{OklopcicHirata2018}. This can additionally alter the predicted magnitude of the absorption signal.

We investigate the effect of a lower XUV flux on the predicted excess absorption signals by reducing the XUV flux from the proxy star by a factor of five, ten and fifteen relative to our original model. We find that decreasing the stellar XUV luminosity for planets b and c results in an initially steep decrease in the outflow temperature, which only decreases marginally with further XUV flux reductions (see Figure~\ref{fig:MCMC}). Both planets see a similar, although less severe effect in the outflow rates, which undergo relatively modest reductions with decreasing XUV flux. These changes in outflow temperature and mass loss rate results in a modest increase in the excess absorption spectra of TOI-4010~b and minimal reductions in the absorption strength for TOI-4010~c (see Figure~\ref{fig:Keck_residuals}). 

For planet d, the outflow rate and temperature only marginally decrease with decreasing XUV flux, however the excess absorption decreases significantly. This suggests that TOI-4010~d is likely in the energy-limited regime of mass loss, as a decrease in the energy deposition results in a decrease in the outflow rate and consequently the observed excess absorption \citep[e.g.,][]{MurrayClay2009, OwenAlvarez16, Lampon21}. 
However, for TOI-4010~d, even the lowest excess absorption (2.4\%) is still readily detectable with Keck. Each planets' predicted mass loss rates fall outside the distribution of mass loss rates and outflow temperatures allowed in our retrievals, and we conclude that a lower stellar XUV flux is unlikely to explain our non-detections.

\subsection{Stellar Winds and Planetary Magnetic Fields}

Strong stellar winds and planetary magnetic fields also play an important role in setting the observability of an outflow. Strong stellar winds can redirect an outflow into a cometary-like tail \citep[e.g.,][]{WangDai2021, MacLeodOklopcic2022} and in the most severe cases, drastically reduce the outflow to a fraction of its Hill sphere \citep{CarolanVidotto2020}. 
Published studies show that the effect of the stellar wind on the He$^*$ signal can vary. \citet{MacLeodOklopcic2022} found that a strong stellar wind can lead to an enhancement of the He$^*$ line depth and extended post-egress absorption, as the confinement of the outflow creates an extended tail of high-density material. Alternatively, studies such as \citet{WangDai2021}, find that outflow models with a stellar wind can at most lead to a reduction in the He$^*$ excess depth by a factor of two. Such a reduction would not reproduce the He$^*$ non-detections observed for the TOI-4010 planets, consistent with the similarly-massed GJ 436 system, in which strong stellar winds were unable to reproduce their He$^*$ non-detection \citep{RumenskikhKhodachenko2023}. We therefore conclude that strong stellar winds do not provide a plausible explanation for our non-detections.

Alternatively, strong planetary magnetic fields can reduce the He$^*$ transit depth by forcing the outflow to follow magnetic field lines. \cite{SchreyerOwen2023} showed that an increase in planetary magnetic field strengths from a \SI{0.3}{\gauss} to \SI{10}{\gauss} field can reduce the excess absorption at most by a factor of three, although we note that the excess absorption does not decrease monotonically with magnetic field strength. If all three planets have surface magnetic field strengths similar to Jupiter, such a reduction in signal size is plausible. However, if all three planets have magnetic field strengths representative of Uranus or Neptune, which are $<5\%$ the surface planetary magnetic field strength of Jupiter \citep[e.g.,][]{2013pss3.book..251B}, the corresponding reduction in excess absorption strength is likely to be much less than a factor of three. 
In the event of a strong Jupiter-like planetary magnetic field, a reduction of this magnitude could only produce a non-detection for TOI-4010~b. Both TOI-4010~c and TOI-4010~d have much larger predicted excess absorption signals that should still be readily detectable even with significant magnetic confinement. 

\begin{figure*}[]
    \centering
    \includegraphics[]{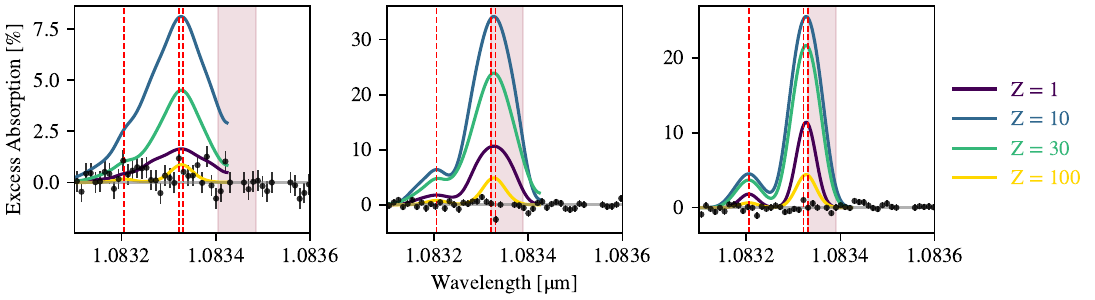}
    \caption{Average Keck/NIRSPEC excess absorption spectra in percent for TOI-4010~b (first panel), TOI-4010~c (second panel) and TOI-4010~d (third panel) shown as black points. Red shaded regions mark the location of a telluric OH line that we masked in our analysis (see Fig.~\ref{fig:Tellurics} and \ref{fig:Keck_residuals}). \texttt{pyTPCI} model predictions for the He$^*$ excess absorption signal as a function of planetary atmospheric metallicity are shown as purple (Z=1), blue (Z=10), green (Z=30), and yellow (Z=100) lines. }
    \label{fig:pyTPCI}
\end{figure*}

\subsection{Enhanced Atmospheric Metallicity}\label{sec:metals}

Up to this point we have assumed a solar metallicity for the planetary outflows in our models. However, increasing the atmospheric metallicity to values greater than $100\times$ solar can drastically reduce the observed He$^*$ signature and overall mass loss rate \citep{ZhangKnutson2022,LinssenShih2024,ZhangBean2025}. As the atmospheric metallicity increases, heating in the atmosphere becomes dominated by the photoionization of metals and metal lines become one of the dominant sources of cooling. This efficient cooling by metal lines reduces the outflow temperature and corresponding mass loss rate. The decreased mass loss rate also results in increased ionization of the outflow, which in turn increases the population of doubly ionized helium. This reduces the fractional He$^*$ population in the outflow, as the primary pathway for producing this state is the recombination of singly ionized helium with an electron. With a reduced population of singly ionized helium, the population of He$^*$ is also reduced \citep{ZhangKnutson2022, ZhangBean2025}. The number density of free electrons available for recombination is also limited at high metallicities. Assuming metal particles are not completely ionized (which is not necessarily true high in the atmosphere, but He$^*$ is formed quite close to the planet), a metal particle typically creates fewer free electrons per unit mass than hydrogen and helium (assuming a non-ionized metal which is not always the case in the upper atmosphere). At high metallicities, the increased metal content will therefore result in a reduction in the free electron density, limiting the amount of He$^*$ that can be produced via radiative recombination \citep{LinssenShih2024}.  

We quantify the effect of increased atmospheric metallicity on the observed He$^*$ signal for the TOI-4010 system by using the same \texttt{pyTPCI} framework from Section~\ref{sec:ares_models} to model outflows with atmospheric metallicities of 10$\times$, 30$\times$, and 100$\times$ solar. Although we also attempted to model outflows with 200$\times$ solar metallicities, these models failed to converge at the onset of advection for all three planets. Following \citet{ZhangBean2025}, we set the abundance of the metals (N, O, Ne, Mg, Si, S, Fe, C) equal to the solar abundance times the atmospheric metallicity. The resulting He$^*$ signals are convolved with the resolution of Keck/NIRSPEC and shown in Figure~\ref{fig:pyTPCI}. 

For all planets, increasing the atmospheric metallicity from solar to 10$\times$ and 30$\times$ solar results in an increase in the predicted He$^*$ signal. However, as the atmospheric metallicity is increased from 30$\times$ to 100$\times$ solar, the 
He$^*$ signal drops below solar predictions. This reduction is sufficient to explain our non-detection for TOI-4010~b, but still results in readily detectable He$^*$ signals for both TOI-4010~c (excess absorption: 4.9\%) and TOI-4010~d (excess absorption: 4.4\%). \citet{ZhangBean2025} found that an increase in atmospheric metallicity from 100$\times$ to 200$\times$ solar can result in an order of magnitude reduction in the predicted He$^*$ excess absorption, suggesting that the non-detections for planets c and d might be explained by a modest additional increase in atmospheric metallicity. TOI-4010 has a metallicity of [Fe/H] = $0.37\pm0.07$~dex, making it one of the most metal-rich hosts of sub-Saturn planets \citep{KunimotoVanderburg2023}. It is therefore reasonable to expect that the atmospheric metallicities of the Neptune-sized planets in this system might be higher than those of Neptunes orbiting stars with lower metallicities. 

\begin{figure*}[]
    \centering
    \includegraphics[]{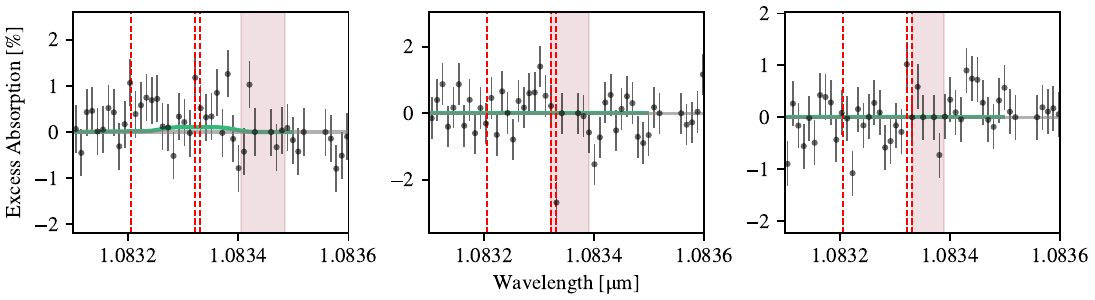}
    \caption{Keck/NIRSPEC average excess absorption spectra in percent for TOI-4010~b (first panel), TOI-4010~c (second panel) and TOI-4010~d (third panel) shown as black points. The shaded red region marks the location of a telluric OH line masked in our analysis (see Fig.~\ref{fig:Tellurics} and \ref{fig:Keck_residuals}). \texttt{AIOLOS} model predictions for the He$^*$ excess absorption signal are shown as green lines. }
    \label{fig:aiolosmodells}
\end{figure*}

\subsection{H/He Fractionation}\label{sec:frac}

Mass-dependent fractionation of the outflow could also act to suppress the He$^*$ transit absorption signal. In this process heavier components, such as helium, decouple from the hydrogen in the uppermost layers of the atmosphere. As a consequence, the outflow can be preferentially depleted in helium, while the atmosphere becomes more enriched in helium over time \citep{zahnlekasting1986, hunten1987}. This process has occurred in the solar system planets \citep{lammer2008} and has likely influenced the transition of early Earth from a hydrogen-rich to a metal-rich atmosphere \citep{zahnle2019}. It is therefore reasonable to investigate its potential impact on the TOI-4010 planets. Indeed, fractionation has already been invoked to predict the existence of He-rich sub-Neptune-sized exoplanets \citep{Malsky2023, Cherubim2024}. The methodologies used by the latter authors differ from the simulations we have presented up to this point, which assumed no fractionation. 

The exact quantity of helium that is dragged into the outflow generally depends on the temperature profile, the mass-flux of hydrogen, the ionization state of helium, and the `injection radius' of helium relative to the wind launching radius.  \citet{schulik2024} have shown that for gas giant exoplanets, outflows can be significantly fractionated at semi-major axis distances greater than $\sim$0.05~au for K stars.  TOI-4010 is also a K star and the outer two planets have semi-major axes within this range \cite[$a=0.06$ and 0.11 au, respectively;][]{KunimotoVanderburg2023},
suggesting that fractionation may play a role in the observed outflows.

In order to estimate the 
expected depletion in helium for the TOI-4010 planets, we perform simple 1D simulations of this problem using the \texttt{AIOLOS} code \citep{schulik2023} including the dissociation of a molecular $\mathrm{H}_2$ layer and molecular cooling from the tables by \citet{malygin2014}. For our simulations, we set the lower boundary radius of the simulation to the $10 \rm~m bar$ level, while injecting helium at a homopause pressure of  $1 \rm~nbar$. The latter value corresponds to the maximum altitude from the solar system planets \citep{atreya1991chapter, catlingkasting2017book}. Overall, this setup allows us to compute the maximum expected helium abundance in a simulation in which the hydrogen is sourced from a molecular layer.
In these simulations we measure the ``reduction factor'', which describes how reduced the helium abundance is in the outflow. In particular, this reduction factor measures the ratio of the mass-loss rates at the sonic point ($\dot{M_{\mathrm{He}}}/\dot{M_{\mathrm{H}}}$, where H and He represent the sum of neutral and ionized hydrogen and helium respectively) normalized to the composition of the atmospheric base ($\rho_{\mathrm{He}}/\rho_{\mathrm{H}}$), which should be approximately representative of the bulk mass of the atmosphere. Normalizing the mass-loss rates results in a reduction factor that measures the composition of the outflow relative to the composition of the atmospheric base (approximately the composition of the bulk atmosphere). Reduction factors equal to one indicate that the helium composition in the outflow is not reduced (fractionation is not occurring) and therefore the compositions of the flow and atmospheric base are equal. For reduction factors less than one, fractionation is occurring, with the severity of fractionation increasing with a decreasing reduction factor (in other words more fractionation reduces the helium present in the outflow).

Our simulation results predict heavily fractionated outflows for planets c and d, 
with reduction factors of $10^{-2}$ for TOI-4010~c and $10^{-6}$ for TOI-4010~d. We additionally find that molecular cooling effectively shuts down the outflow for planet d, as shown in Figure~\ref{fig:aiolosmodells}. This effect is not seen in our previous \texttt{pyTPCI} simulations, as those models do not include molecular line cooling.
TOI-4010~b is a special case, as its outflow lies in the photon-limited regime \citep{OwenAlvarez16}, where the stellar heating distributes itself over the entire radial extent of the flow. This leads to 
a rapid mass loss rate ($\sim 20$~km~s$^{-1}$ at 10 $R_p$) and near-perfect coupling of helium into the outflow, resulting in a reduction factor of $1$. However, the helium in the outflow has insufficient time to ionize and remains mostly neutral, with the ion fraction reaching only $5\%$ at 10 $R_p$, reducing the strength of the predicted He$^*$ absorption signal during transit as presented in Fig.~\ref{fig:aiolosmodells}.

These simulations result in nearly complete reductions in the predicted excess absorption signals for TOI-4010~b, c, and d relative to the solar metallicity \texttt{pyTPCI} model predictions. These reductions are consistent with our non-detections, and we conclude that fractionation and/or molecular cooling can indeed suppress the detection of He$^*$ signals for this system. We note that the dominant source of cooling in the \texttt{AIOLOS} models is the molecular cooling and Ly$\alpha$ cooling for the atomic region, which differs from the dominant source of cooling (bremsstrahlung or free-free radiation cooling) in the \texttt{pyTPCI} models. These differences explain the difference in the predicted mass loss rates, as cooling mechanisms and the mean molecular weight control the outflow rate, which in turn determines the ionic state of the elements in the outflow. More detailed models that include both fractionation and tunable atmospheric metallicity are needed in order to study the competing effects of these variables on outflow strengths for these planets.

\subsection{Constraints on Past Evaporation Histories}\label{sec:evo}

In the preceding subsections, we have evaluated possible mechanisms that would suppress signatures of present-day mass loss. In this subsection, we quantify what our non-detections of present-day mass loss mean for the past mass loss histories of all three planets.
In order to study the past evaporation of these planets and place constraints on their internal structures, we performed simulations of their past evaporation histories taking into account the XUV emission history of their host star.
To do so, we adopted the method described in \citet{Fernandez2023:photoevolver}, which combines three models: (1) a description of the XUV emission history of the host star, (2) a model of the internal structure of the planet (including the thermal evolution of its gaseous envelope), and (3) a model of atmospheric escape to compute the mass loss rate from the planet at each point in time.

\begin{figure*}[]
    \centering
    \includegraphics[width=\textwidth]{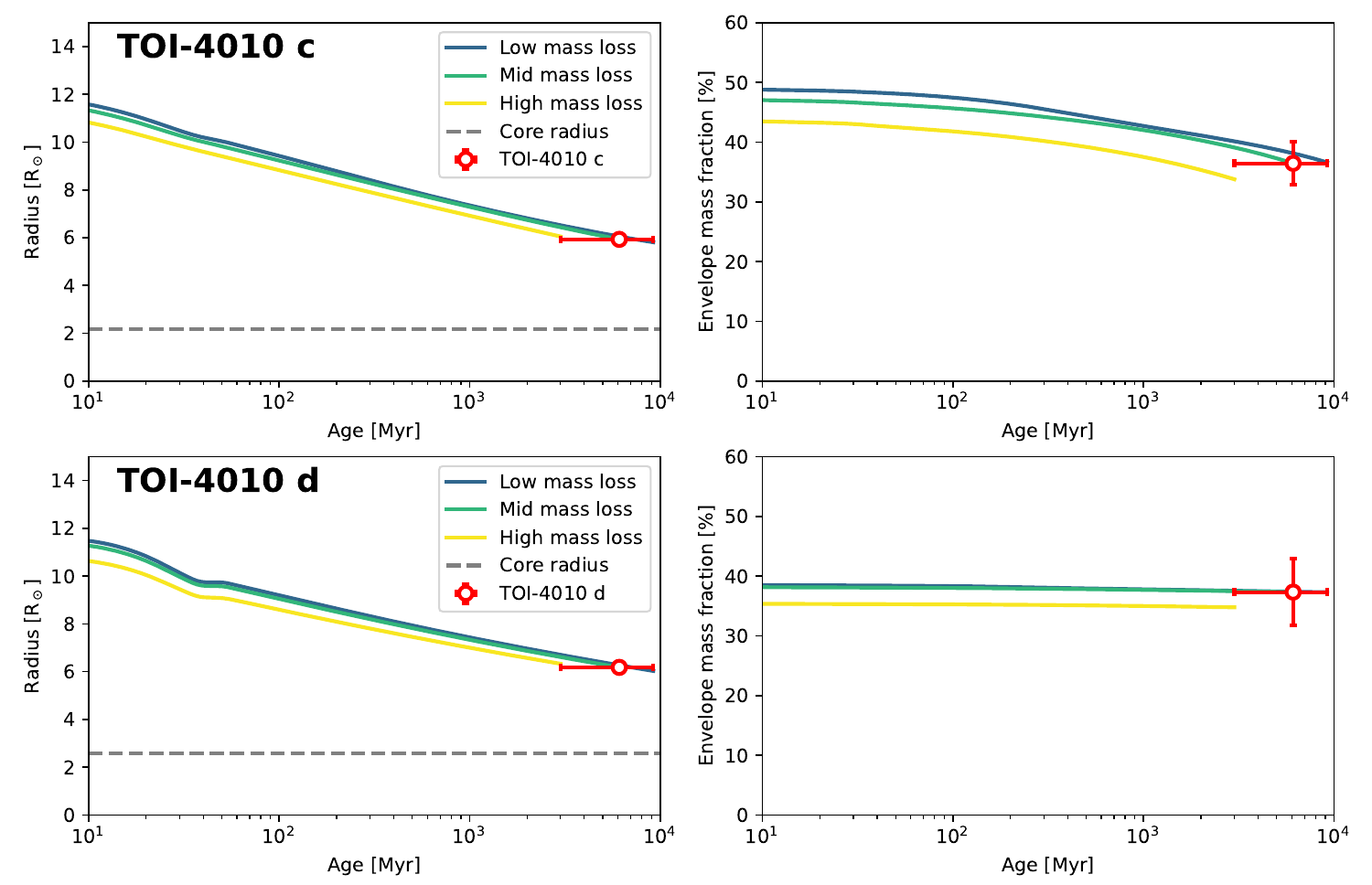}
    \caption{
    Top panels: Evolution of the radius (left panel) and envelope mass fraction (right panel) for TOI-4010\,c, as described in Section\,\ref{sec:evo}. The solid lines show atmospheric evolution under different XUV irradiation histories and different values for the planet's parameters within their $1\sigma$ uncertainty. The planet's present-day location on each panel is plotted as a red circle.
    Bottom panels: Evolution of the radius and envelope mass fraction for TOI-4010\,d, following the top panels.
    }
    \label{fig:EvoPlanetsCD}
\end{figure*}

\begin{figure*}[]
    \centering
    \includegraphics[width=\textwidth]{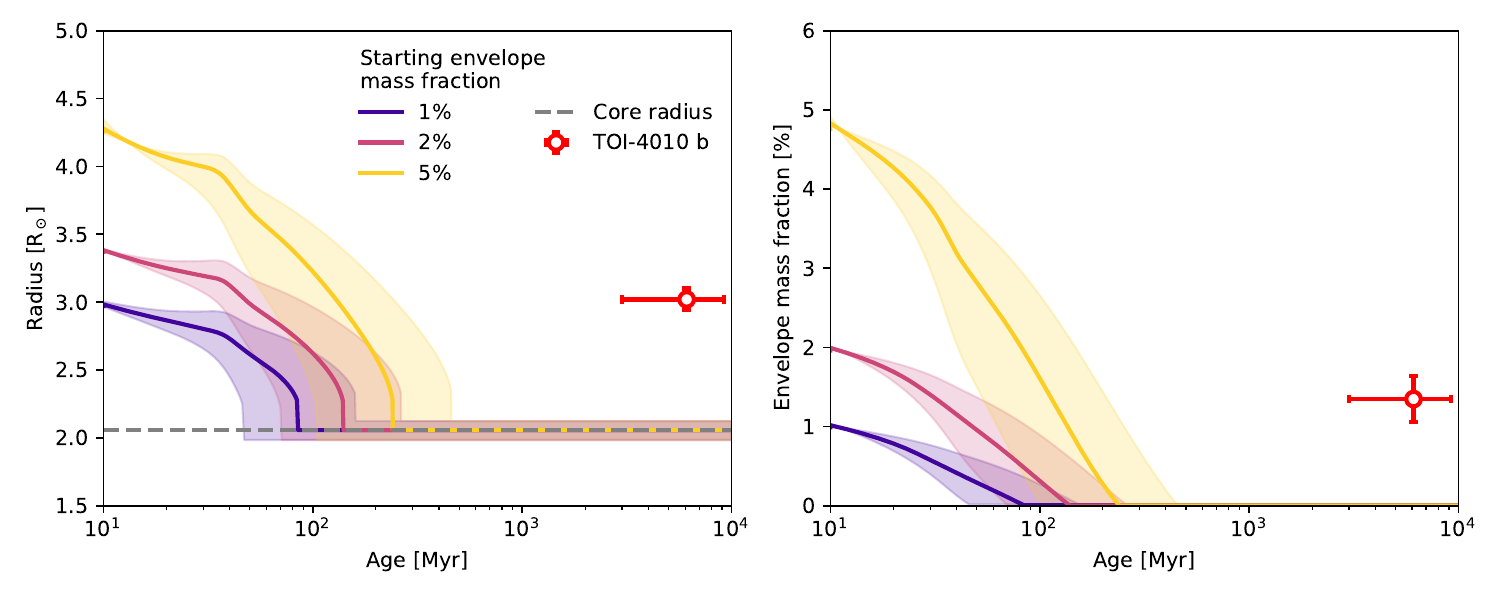}
    \caption{
    Left panel: Evolution of the radius of TOI-4010\,b under atmospheric escape and thermal contraction, as described in Section\,\ref{sec:evo}. The solid lines show evolution under different starting envelope mass fractions of 1\%, 2\%, and 5\%, and the shaded regions represent the uncertainties on each evaporation history, given the uncertainties in the planet's parameters and the XUV emission history of its host star. The planet's present-day age and radius are plotted as a red circle.
    Right panel: Evolution of the envelope mass fraction of the gaseous atmosphere of TOI-4010\,b, following the left panel.
    }
    \label{fig:EvoPlanetB}
\end{figure*}

To model the past XUV emission of TOI-4010, we adopted the rotational and XUV stellar evolution models by \citet{JohnstoneBartel2021}. Rotational evolution models serve as a proxy for the XUV emission history as spin period and stellar X-ray activity are linked via the rotation-activity relation, where faster rotators are more X-ray active \citep[e.g.][]{Pizzolato2003:rotation-activity, Wright2011:rotation-activity}.
We find that the measured age and spin period of TOI-4010 \citep[age: $6.1\pm3.1$~Gyr; spin period: 37.7~days;][]{KunimotoVanderburg2023} is consistent with the \citet{JohnstoneBartel2021} rotation models, which predict spin periods between 23 and 51\,days at ages between 3 and 9\,Gyr.
However, we cannot determine whether the star was a fast or slow rotator when it was younger, and thus whether its earlier X-ray emission was bright or faint. This is because the diversity of spin periods that young stars exhibit is erased at later ages as stars spin down and converge into a single mass-dependent relation \citep{Johnstone2015:stellar-wind}.
For that reason, we considered both bright and faint XUV emission histories for the host star in our simulations of the planets' evaporation histories -- these correspond to the 5th and 95th percentile XUV emission models from \citet{JohnstoneBartel2021}, which span the $2\sigma$ spread in rotation periods observed in stellar populations.

Next, to model the internal structures of the planets, we assumed a two-layer structure with a solid core surrounded by a H/He solar metallicity envelope. We note that our constraints on present-day mass loss rates are lower than the mass loss rates we predict for solar metallicity envelopes, indicating the atmospheric composition might deviate from solar (see Section~\ref{sec:metals} \& \ref{sec:frac}). However, we use the assumption of solar metallicity envelopes as an upper bound on the past evaporation histories of these three planets. For the solid core, we adopted the empirical mass-radius relation for rocky cores by \citet{Otegi2019RevisitedMasses}, and for the gaseous atmosphere we adopted the envelope structure model by \citet{Chen2016:atm-model} based on MESA simulations.
Moreover, we adopted the mass loss model by \citet{Kubyshkina2018:mass-loss-model}, based on hydrodynamic simulations of escaping atmospheres; their model accounts for both X-ray and EUV driven escape as well as multiple regimes of mass loss, including energy-limited and recombination-limited photoevaporation as well as core-powered mass loss.

Finally, we adopted the {\tt photoevolver} code \citep{Fernandez2023:photoevolver} to combine these models and simulate the evaporation past of the three planets.
We used the RK45 algorithm as the integration method with a variable time step no larger than 1~Myr.
Additionally, we explored a diversity of evaporation histories for each planet in order to account for the uncertainties in their radii, masses, and ages, as well as the unknown XUV emission history of the host star.

For TOI-4010\,c and d, we evolved their current states backwards in time from the present day to an age of 10\,Myr, corresponding to the estimated time when the protoplanetary disk has dispersed and boil-off processes have stopped \citep{RogersOwen2024}.
Figure~\ref{fig:EvoPlanetsCD} shows the results of our simulations, which suggest that both TOI-4010\,c and d have lost relatively little of their overall H/He envelope mass to atmospheric escape.
Figure~\ref{fig:EvoPlanetsCD} also shows that the initial radii of both planets were 3\,R$_\oplus$ larger than their present-day values at an age of 10\,Myr; their radii decrease over time as a result of radiative cooling \citep{Chen2016:atm-model}.

For TOI-4010\,b we considered initial gas envelopes with mass fractions of 1\%, 2\%, and 5\%, and evolved them from the age of 10\,Myr to the present day. We do not consider larger starting envelopes, as they would have been stripped by boil-off processes during the dispersal of the protoplanetary disk \citep[e.g.][]{RogersOwen2024}. Our results, shown in Figure~\ref{fig:EvoPlanetB}, suggest that all scenarios result in the complete loss of this initial gas envelope within the first 500\,Myr, much earlier than the current age of the planet. Given that TOI-4010~b has a present-day bulk density consistent with a H/He envelope \citep{KunimotoVanderburg2023}, it is clear that its time-integrated mass loss rate must be significantly lower than predicted by these models. This is consistent with our previous conclusion that either a high atmospheric metallicity (at least 100$\times$ solar) or low He$^*$ abundance are needed to explain our non-detection of a He$^*$ absorption signal during the transit of TOI-4010~b.  The former would result in a globally reduced mass loss rate, making it easier for TOI-4010~b to retain a hydrogen-rich envelope.  

Using our model framework we also estimate present-day mass loss rates of all three planets. Using the present-day radii and masses for the TOI-4010 planets, along with the mass-loss model of \citet{Kubyshkina2018:mass-loss-model} and adopting a predicted XUV flux of $(3.0\pm1.0)\times10^{28}$~erg~s$^-1$ (uncertainties propagated from errors in stellar age and rotation period, we note that this estimate agrees with the XUV flux of HD 85512: $1.2\times10^{28}$~erg~s$^-1$) from the rotation-activity models of \citet{JohnstoneBartel2021}, we calculate present-day mass-loss rates (assuming the upper atmosphere is pure H/He) of $10^{11.00\pm0.16}$~g~s$^{-1}$, $10^{10.65\pm0.15}$~g~s$^{-1}$ and $10^{9.73\pm0.12}$~g~s$^{-1}$ for TOI-4010~b, c and d respectively. The estimated present-day mass loss rate for TOI-4010~c is slightly higher than our 95\% upper bound ($10^{10.53}$~g~s$^{-1}$) from Section~\ref{sec:masslosssub}, but still within $1\sigma$, whereas our estimated mass loss rate for TOI-4010~d is consistent with our upper bound ($10^{10.50}$~g~s$^{-1}$). However, the present-day mass loss rate for TOI-4010~b is much higher than our estimated 95\% upper bound ($10^{10.17}$~g~s$^{-1}$). This discrepancy indicates that there must be some method of mass loss suppression not accounted for in these solar metallicity H/He models that could reduce the predicted mass-loss to a value consistent with our 95\% upper bound and non-detection, such as an enhanced atmospheric metallicity (see Section~\ref{sec:metals}) or a reduced He$^*$ abundance, although we note that the latter would only suppress the detectable signal and not the outflow itself (see Section~\ref{sec:frac}).

\section{Conclusion}\label{sec:Conclusion}
In this work, we use Keck/NIRSPEC to search for evidence of He$^*$ absorption in the transmission spectra of all three planets in the TOI-4010 system, and find that none have detectable He$^*$ absorption signals. We place an upper limit on the possible He$^*$ absorption for each planet by averaging the excess absorption within \SI{0.75}{\angstrom} of the main helium triplet peak, which allows us to rule out absorption to 1.23\%, 0.81\%, and 0.87\% for TOI-4010~b, c, and d respectively at 95\% confidence. Using our transmission spectra and Parker wind models, we obtain upper limits on the possible mass loss rates for each planet of: $10^{10.17}$g~s$^{-1}$, $10^{10.53}$g~s$^{-1}$, and $10^{10.50}$g~s$^{-1}$ for TOI-4010~b, c, and d respectively.

One-dimensional outflow models assuming a solar composition predict that all three planets in the TOI-4010 system should have easily detectable He$^*$ excess absorption. We attempt to explain our non-detections by considering four potential factors that could weaken or suppress the outflow: a decreased host star XUV luminosity, stellar winds or planetary magnetic fields, enhanced planetary atmospheric metallicities, and fractionation. We find that a decreased host star XUV luminosity, as well as strong stellar winds and planetary magnetic fields, are not individually sufficient to reduce the 1D outflow predictions to a level that is consistent with our non-detections. However, it is possible that a combination of these factors could explain our non-detections.  This could be more rigorously quantified in future studies by developing mass loss models that incorporate tunable stellar wind strengths, XUV luminosities and planetary magnetic field strengths.

For TOI-4010~b, an atmospheric metallicity of 100$\times$ solar reduces the predicted He$^*$ signal to a level that is consistent with our non-detection. However, our upper limits for TOI-4010~c and d appear to require much larger metallicity enhancements ($\simeq200\times$ solar) to suppress the outflow signals. Given the high stellar metallicity of TOI-4010, it is possible that the atmospheric metallicities of these planets could be enhanced relative to those of other Neptune-sized planets. This can be tested by upcoming JWST observations of the TOI-4010 system, which should be able to constrain the atmospheric metallicities of all three planets.

Alternatively, solar metallicity models including helium fractionation in the outflow and/or molecular cooling and other radiative effects can reproduce our non-detections for all three planets. 
It is also possible that both atmospheric metallicity and these other effects work in tandem to reduce the expected outflow strength on the TOI-4010 planets. Studies of hydrodynamic escape at non-solar metallicities show that molecular and atomic cooling can suppress the mass-loss rate \citep{Yoshida2020, Yoshida2022, Yoshida2024, Nakayama2022}, with this suppression increasing with increasing atmospheric metallicity \citep{ZhangKnutson2022, LinssenShih2024, ZhangBean2025}. This mass loss suppression will slow down the outflow, rendering it more prone to fractionation. Therefore, it is plausible that the combined effects of atmospheric metallicity and fractionation could also suppress the outflows on the TOI-4010 planets. Future mass loss models that incorporate the effects of tunable atmospheric metallicity and fractionation are therefore paramount to understanding the complex outflow properties  and behavior on Neptune and sub-Neptune-sized worlds. We note that while our calculations show that it is plausible that fractionation suppresses the visible He$^*$ signal for the TOI-4010 system, it is currently unclear whether this applies to the population of He$^*$ detections and non-detections as a whole, see e.g., the compendia by \cite{dosSantosreview} and \cite{orellmiquel2024}. Those authors particularly note a strong dependency on stellar type for the non-detections, with planets hosted by M stars having the largest frequency of non-detections. Future studies should explore whether fractionation can be applied universally to explain this trend.

We also evaluate the past evaporation histories for all three planets and find that TOI-4010~c and d have lost relatively little of their initial H/He envelope while TOI-4010~b should have been completely stripped of it's envelope within the first $\sim$500~Myr. However, TOI-4010~b's current bulk density is consistent with a puffy, H/He envelope, indicating a source of outflow suppression that must persist over the planet's lifetime. Our modeling efforts indicate that such an outflow suppression can only be achieved by high atmospheric metallicities (100$\times$ solar), as a low He$^*$ abundance would only suppress the helium outflow signal and not the overall mass loss rate. 
The presence of a significant gas envelope on TOI-4010~b, as well as the low predicted percentage of past envelope loss on TOI-4010~c and d, place limits on the degree to which these planets may have been affected by photoevaporative mass loss. Our observations indicate that the higher envelope mass fraction of planet d relative to planet c is likely primordial, but our constraints for planet b are weaker and allow for a correspondingly wide range of initial envelope mass fractions.

\section*{acknowledgments}

This research has made use of the NASA Exoplanet Archive, which is operated by the California Institute of Technology, under contract with the National Aeronautics and Space Administration under the Exoplanet Exploration Program. This research also made use of \texttt{nep-des} (available in \url{https://github.com/castro-gzlz/nep-des}). This material is based upon work supported by the National Science Foundation Graduate Research Fellowship Program under Grant No.~DGE‐1745301. Any opinions, findings, and conclusions or recommendations expressed in this material are those of the author(s) and do not necessarily reflect the views of the National Science Foundation. 

We thank the Keck Observatory telescope and support astronomers, with special thanks to Greg Doppmann, Percy Gomez, Anthony Connors, Arina Rostopchina, and John Pelletier. We thank Justin Moore for his significant help in debugging software issues.

Co-author M. Zhang thanks the Heising-Simons Foundation for funding his 51 Pegasi b postdoctoral fellowship. Co-author J. Fernández Fernández acknowledges studentship support from the UK Science and Technology Facilities Council (STFC) and co-author P.J. Wheatley acknowledges STFC support under consolidated grants  ST/T000406/1 and ST/X001121/1.


\vspace{5mm}
\facilities{ADS, NASA Exoplanet Archive, Keck II/NIRSPEC}

\software{\texttt{AIOLOS} \citep{schulik2023}, \texttt{arviz} \citep{KumarCarroll2019}, \texttt{astropy} \citep{AstropyCollaborationRobitaille2013, AstropyCollaborationPrice-Whelan2018, AstropyCollaborationPrice-Whelan2022},  \texttt{emcee} \citep{Foreman-MackeyHogg2013}, \texttt{matplotlib} \citep{Hunter2007},
          \texttt{numpy} \citep{HarrisMillman2020}, \texttt{photoevolver} \citep{Fernandez2023:photoevolver},   \texttt{p-winds} \citep{DosSantosVidotto2022}, \texttt{pyTPCI} \citep{RosenerZhang2025}, \texttt{scipy} \citep{VirtanenGommers2020}
          }

\bibliography{references}{}
\bibliographystyle{aasjournal}

\end{document}